Review Article

# The Mitchell Spectrograph: Studying Nearby Galaxies with the VIRUS Prototype

**Guillermo A. Blanc**

*Observatories of the Carnegie Institution for Science, 813 Santa Barbara Street, Pasadena, CA 91101, USA*

Correspondence should be addressed to Guillermo A. Blanc; gblancm@obs.carnegiescience.edu

The Mitchell Spectrograph (a.k.a. VIRUS-P) on the 2.7 m Harlan J. Smith telescope at McDonald Observatory is currently the largest field of view (FOV) integral field unit (IFU) spectrograph in the world ($1.7' \times 1.7'$). It was designed as a prototype for the highly replicable VIRUS spectrograph which consists of a mosaic of IFUs spread over a $16'$ diameter FOV feeding 150 spectrographs similar to the Mitchell. VIRUS will be deployed on the 9.2 meter Hobby-Eberly Telescope (HET) and will be used to conduct the HET Dark Energy Experiment (HETDEX). Since seeing first light in 2007 the Mitchell Spectrograph has been widely used, among other things, to study nearby galaxies in the local universe where their internal structure and the spatial distribution of different physical parameters can be studied in great detail. These observations have provided important insight into many aspects of the physics behind the formation and evolution of galaxies and have boosted the scientific impact of the 2.7 meter telescope enormously. Here I review the contributions of the Mitchell Spectrograph to the study of nearby galaxies, from the investigation the spatial distribution of dark matter and the properties of supermassive black holes, to the studies of the process of star formation and the chemical composition of stars and gas in the ISM, which provide important information regarding the formation and evolution of these systems. I highlight the fact that wide field integral field spectrographs on small and medium size telescopes can be powerful cost effective tools to study the astrophysics of galaxies. Finally I briefly discuss the potential of HETDEX for conducting studies on nearby galaxies. The survey parameters make it complimentary and competitive to ongoing and future surveys like SAMI and MANGA.

## 1. Introduction

The Mitchell Spectrograph (a.k.a. VIRUS-P) was commissioned in early 2007 on the 2.7 m Harlan J. Smith telescope at McDonald Observatory. It is owned and operated by the University of Texas at Austin and was designed as a prototype for the Visible Integral Field Replicable Unit Spectrograph (VIRUS; [1]), a massively replicated fiber-fed IFU spectrograph for the 9.2 m Hobby-Eberly Telescope (HET). VIRUS consists of a mosaic of 75 $50'' \times 50''$ IFUs spread regularly with a 1/4 filling factor over a $16'$ diameter field of view. The IFUs themselves have a 1/3 filling factor and feed 150 spectrographs similar to the Mitchell Spectrograph. VIRUS will be used to conduct the HET Dark Energy Experiment (HETDEX; [2–4]), a large ($\sim$90 deg$^2$) blind integral field spectroscopic survey aimed at studying the properties of dark energy at high redshift by measuring the power spectrum of $\sim 8 \times 10^5$ Ly$\alpha$ emitting galaxies (LAEs) at $1.9 < z < 3.5$. VIRUS is currently under construction and first light is expected to occur in mid-2014.

In its 6-year lifetime the Mitchell Spectrograph has provided a test bench for the technology and the components involved in the construction of VIRUS and a proof of concept for the observing strategy, data processing, and analysis algorithms that will be used in HETDEX [3]. Although these were the main goals behind the design and construction of the Mitchell Spectrograph, the prototype has proved to be a unique and powerful instrument for the study of galaxies not only in the nearby universe but also at high redshift (e.g., [3–7]).

In this paper I review the impact that the Mitchell Spectrograph has had on the field of galaxy astrophysics. Numerous studies that use Mitchell Spectrograph data have been published during the last few years. These studies are focused on a large range of astrophysical processes that are important for galaxy evolution, including the physics of super massive black holes (SMBHs) in the centers of galaxies [8–11]; the spatial structure of the dark matter (DM) halos in which galaxies live [12, 13]; the chemical composition, dynamics, and physical conditions of gas in the interstellar medium (ISM) and the circumgalactic medium (CGM) of galaxies [14–16]; the process of star formation on galactic scales [17, 18]; and the physical properties of the stellar populations present in different regions of nearby star forming and early-type systems [19–21].

During the last decade a new generation of wide field integral field spectrographs, like SAURON on the 4.2 m William Herschel Telescope [22], PPAK on the 3.5 m at Calar Alto Observatory [23], SparsePak on the WIYN 3.5 m telescope [24], and the Mitchell Spectrograph on the 2.7 m, have demonstrated the power of integral field spectroscopy as a tool that is extremely well suited for small and medium aperture telescopes. These instruments have been systematically used to study nearby galaxies. Surveys like SAURON [25], ATLAS3D [26], PINGS [27], CALIFA [28], VENGA [29], and VIXENS [18] either have or are in the process of building revolutionary spectroscopic datasets on these systems.

The large fields of view and large plate scales that typically characterize these instruments allow for an efficient mapping of large areas on the sky and make these instruments extremely sensitive to low surface brightness emission. At a fixed focal ratio, the plate scale is inversely proportional to the telescope aperture size. Therefore, modulo differences in transmission between different telescopes, the surface brightness sensitivity of an IFU with a fixed spatial resolution element size in physical units is independent of the size of the telescope in which it is installed. This fact, combined with moderate instrument building costs and the relative facility with which large amounts of time can be scheduled for an observing program in smaller facilities, makes wide field IFU spectrographs extremely cost effective and scientifically competitive instruments for small and medium size telescopes. I hope this review helps spread this idea by presenting the Mitchell Spectrograph as an example of what can be achieved with this type of instruments.

In Section 2 I provide a brief description of the instrument and its various components. I present a showcase of scientific results on nearby galaxies and discuss ongoing surveys being conducted with the Mitchell Spectrograph in Section 3. I discuss the prospects for studying nearby galaxies with VIRUS in the context of the HETDEX survey in Section 4 and present some final conclusions in Section 5.

## 2. The Mitchell Spectrograph (a.k.a. VIRUS-P)

The Mitchell Spectrograph (http://www.as.utexas.edu/mcdonald/facilities/2.7m/virus-p.html) design and construction are described in detail in Hill et al. [30]. The instrument is shown in Figure 1. The IFU uses a densepak-type 246 fiber bundle. Fibers are arranged following an hexagonal pattern with a 1/3 filling factor across a square $1.7' \times 1.7'$ FOV, requiring three dithered observations to ensure an almost complete coverage of the field. The IFU is fed at f/3.65 through a telecentric, two-group dioptric focal reducer. At the plate scale of the 2.7 m telescope fibers subtend a $4.2''$ diameter on sky.

The spectrograph has a double-Schmidt optical design. It uses a volume phase holographic (VPH) grating at the pupil between the articulating f/3.32 folded collimator and the f/1.33 cryogenic prime focus camera. Gratings are interchangeable and can be blazed at different angles. A set of four VPH gratings are available which can be used in different modes to provide spectral resolutions in the $800 < R < 4000$ range and sample different bandpasses across the 3600–6900 Å range. High on-sky throughput is achieved by the use of high reflectivity dielectric coatings. The Mitchell Spectrograph is gimbal-mounted on the telescope to allow short fibers for high UV throughput, while maintaining high mechanical stability. This translates in good instrument stability and observing efficiency as it eliminates the need for calibration frames at the position of the science targets. The instrument software and the $4.5' \times 4.5'$ field, fixed-offset guider provide rapid acquisition, guiding, and precision dithering. A custom data reduction pipeline called VACCINE [3] yields Poisson noise limited sky subtracted spectra.

## 3. Nearby Galaxies Studies

Thanks to the fact that we can resolve nearby galaxies down to small physical scales, they offer unique laboratories in which we can test our understanding of the critical processes involved in galaxy formation and evolution. Integral field spectroscopy allows one to study important processes affecting baryons in galaxies like star formation, chemical enrichment, and feedback, in a spatially resolved manner. Furthermore, wide field stellar and gas kinematics can be used to constrain the shape of the dark matter gravitational potential in which all these processes take place. Being able to map the kinematics and physical properties of stars and gas across the different environments and morphological structures present in galaxies allows one to study in detail the fossil record of the processes by which these objects have formed and evolved.

In this section I provide an overview of the contributions of the Mitchell Spectrograph to the study of nearby galaxies. This might not be a complete compilation of papers in the literature that use Mitchell data but rather a selection of the most relevant works to the field of nearby galaxy astrophysics that I am aware of and that are based on Mitchell Spectrograph data.

*3.1. Stellar Populations: The Assembly of Nearby Spheroids and Disks.* By modeling the relative strength of different absorption features in the integrated stellar spectrum of galaxies or regions within galaxies we can estimate relevant physical properties of the stellar populations that are present (e.g., [31, 32]). Integral field spectroscopy provides two important advantages at the time of conducting this type of measurement in nearby galaxies. First, the two-dimensional nature of the data allows one to study stellar populations as a function of position or environment within galaxies in a way that is challenging for slit spectroscopy. Second, integral field spectroscopy allows the observer to integrate the spectra over large arbitrarily shaped regions across nearby galaxies. Smart binning schemes can be used to boost the signal to noise (S/N) of the spectra allowing the measurement of stellar population parameters in low surface brightness regions. The latter, combined with the good low surface brightness sensitivity at the native spatial resolution of IFUs with large spaxels like the Mitchell Spectrograph (4.2″ diameter fibers), has opened a new avenue in the study of stellar populations in nearby galaxies by permitting measurements at large galactocentric radii. This is important as, in a hierarchical universe like the one we live in, the outskirts of galaxies can provide valuable information regarding their merging and gas accretion history.

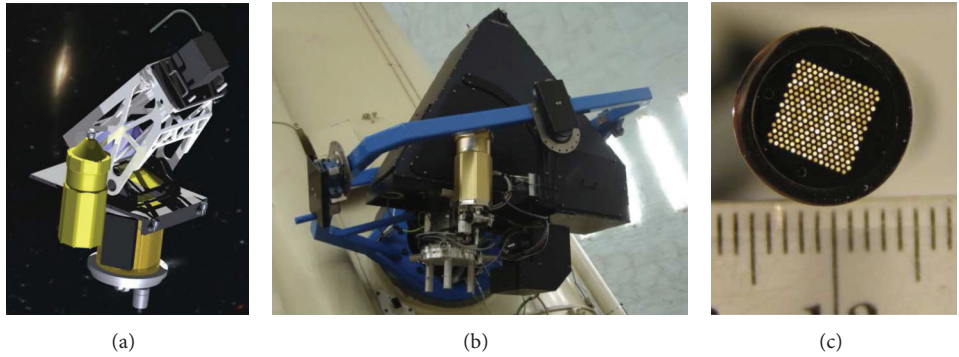

(a) (b) (c)

Figure 1: The Mitchell Spectrograph. (a) Diagram showing the interior of the instrument. The casing of the fiber output pseudoslit can be seen in black at the top right, from there light is reflected onto the collimator mirror (purple), back onto a flat mirror at the position of the pseudo-slit, and down through the VPH grating (yellow) and into the Schmidt camera at the bottom of the spectrograph (gold). Dispersed light is reimaged onto a Fairchild Instruments 2048 × 2048 backside illuminated, antireflection coated CCD. The liquid $N_2$ dewar is also shown in the foreground. (b) The Mitchell Spectrograph installed on the broken Cassegrain focus of the 2.7 m telescope. The gimbal mount can be seen in blue. (c) A zoom into the input head of the 246 fiber IFU.

Almost all aspects of galaxy formation and evolution are encoded in the physical properties of the stellar populations present in galaxies. The metallicities, abundance ratios, and ages of different generations of stars across a galaxy are set by the history of gas accretion, star formation, and feedback induced outflows within the galaxy itself and its accreted satellites. The Mitchell Spectrograph has been used to study stellar populations in both nearby early-type (E, S0) and spiral galaxies.

Elliptical galaxies have long been known to show color gradients with bluer colors towards their outer regions (e.g., [33]). Spectroscopic stellar population studies have shown that these color gradients are mainly driven by gradients in metallicity with the abundance of metals decreasing towards larger radii (e.g., [34, 35]). In view of the recent discovery of a population of massive compact passively evolving galaxies at $z = 1-2$ [36, 37] and the dramatic size evolution implied if one attempts to link these objects as progenitors of present day elliptical galaxies, studying the stellar populations present in the halos of nearby massive ellipticals can shed light on the processes that drive this evolution. The most likely scenario currently invoked to explain the growth of the outer halo of early type galaxies is late time minor merging at $z < 1$ ([38, 39]).

In Greene et al. [21] the authors use the Mitchell Spectrograph to observe a sample of eight nearby (~85 Mpc) early type galaxies (six ellipticals and two S0s) with the goal of conducting a spatially resolved stellar population analysis out to large radii (2.5 $R_e$). In only 40 minutes of effective exposure time per fiber (i.e., per dither) and after coadding the spectra in elliptical annuli with a width of 0.5 $R_e$ (~4″ or one fiber at the typical distance of the targets) they are able to reach S/Ns~40 in continuum for the outermost radial bins at 2.5 $R_e$. Figure 2 shows an example of the azimuthally binned spectra for NGC 7509. Using these high S/N spectra the authors can make quality measurements of the EWs of the H$\beta$, Mg$b$, and ⟨Fe⟩ absorption features out to large galactocentric radii (see Figure 2). These spectral indexes are used to derive gradients in age, metallicity [Fe/H], and $\alpha$-abundance [$\alpha$/Fe].

A common trend in all the galaxies observed in Greene et al. [21] is the presence of a steep gradient in the Mg$b$ EW which drops towards the outer regions while the ⟨Fe⟩ and H$\beta$ indexes remain roughly flat out to 2.5 $R_e$. Modeling the observed EWs the authors find that this average behavior is driven by significant [Fe/H] gradients in the majority of the galaxies accompanied by a subtler gradient in [$\alpha$/Fe] which decreases slightly towards large radii. This is the first detection of an [$\alpha$/Fe] gradient in elliptical galaxies.

One interesting result of this study is the observed change in behavior of the well established Mg$b$-$\sigma_*$ relation [40, 41] when considering regions at larger distance from the galaxy nucleus. Within $R_e$ elliptical galaxies show a tight correlation between the Mg$b$ index and the effective stellar velocity dispersion. Greene et al. [21] find that if one measures Mg$b$ in regions outside $R_e$ the normalization of the Mg$b$-$\sigma_*$ relation decreases, the scatter increases significantly, and there are indications of a steepening. These effects are stronger towards larger radii. They reflect a clear change in the stellar populations of the outskirts of early type galaxies with respect to their centers and are consistent with a scenario in which the stellar halos of elliptical galaxies are formed by the accretion of objects formed at earlier times in small halos that have truncated star formation histories.

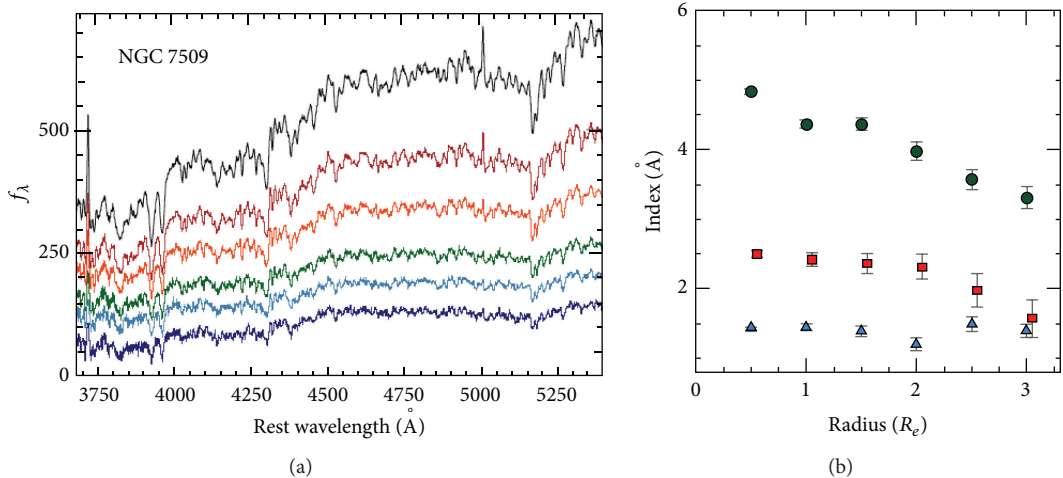

FIGURE 2: (a) Coadded spectra in elliptical annuli at $0 – 0.5\,R_e$, $0.5 – 1\,R_e$, $1.5 – 2\,R_e$, and $2 – 2.5\,R_e$ (from top to bottom) for NGC 7509. (b) Index EW radial profiles for Mg$b$ (green), $\langle Fe \rangle$ (red), and H$\beta$ (blue). Taken from Greene et al. [21].

By matching the observed Mg$b$ EW at $>2\,R_e$ to galaxy mass using the relations of Graves et al. [42] the authors find that the stellar halo of present day ellipticals was most likely formed by the accretion of satellites with stellar masses which are typically a factor of ~10 smaller than the central object. The data also implies that the progenitors of the stellar halo had lower metallicities and higher $\alpha$-abundance than present day ellipticals of similar masses. This is in contrast with predictions from the "monolithic collapse" scenario in which $\alpha$-abundances are lower than expected in the stellar halo as winds can eject metals more easily than from the central regions where the potential is deeper [43]. The authors conclude that *"the outer parts of these galaxies [massive ellipticals] are built up via minor merging with a ratio of ~10 : 1, but that the accreted galaxies did not have sufficient time to lower their $\alpha$-abundance ratios to those seen in ~$L_*$ elliptical galaxies today."*

The Mitchell Spectrograph has also been used to study the stellar populations in the outskirts of late-type star forming disk galaxies by Yoachim et al. [19, 20]. These two papers are aimed at studying the nature of breaks in the exponential surface brightness profiles of disk galaxies at large radii. Pohlen and Trujillo [44] used the Sloan Digital Sky Survey (SDSS) to show that the majority (60%) of late-type disk galaxies in the local universe show truncated exponential profiles which transition to a steeper exponential at some break radius typically found at 1.5–4.5 times the scale length of the inner component. Of the rest of their sample, 30% show shallower exponential profiles outside the break radius, and 10% can be best fit by a single component.

The classical explanation for the existence of a truncation in the edge of disk galaxies is the presence of a star formation threshold [45] or more likely a significant drop in the star formation efficiency in the low gas surface density environments typical of the outskirts of galaxies [46]. Work by Schruba et al. [47] has shown that this low star formation efficiency at large radii is caused by a low-molecular-to-atomic gas fraction and not by an actual drop in the star formation efficiency of molecular gas. On top of in situ star formation, another process that could be important at populating the outskirts of disk galaxies is secular stellar migration caused by gravitational interactions with disk structure (e.g., [48, 49]). Since stellar migration is a random walk stars require longer times to venture further out in the disk. Therefore if the stellar populations beyond the edges of disks are dominated by this process we expect to observe significant changes in stellar population parameters beyond the break radius, where the contrast between migrated and formed in situ stars is the greatest and the ages of migrated stars are the oldest.

In Yoachim et al. [20] the authors observe a sample of twelve nearby spiral galaxies chosen to match the Mitchell Spectrograph field of view. The bulk of the sample (9/12) shows truncated photometric profiles, one object shows an upbending profile, and the remaining two galaxies are best fit by single exponentials. In only 20–45 minutes of effective exposure time and after binning the spectra in $4''$ wide elliptical annuli they obtain spectra deep enough (S/N~ 40) to measure stellar population parameters down to a $V$-band surface brightness 23-24 mag arcsec$^{-2}$. This allows them to successfully measure ages outside the break radius for six galaxies in the sample (all of them with downbending profiles).

An interesting contribution from this work comes from the detailed study that the authors do of the systematics associated measuring stellar population parameters in late-type systems. While individual absorption features (e.g., Lick indices) have been successfully used to study the stellar populations of old passively evolving early-type galaxies (as in the Greene et al. [21] study described above), they are not reliable tools to use for studying star forming objects. The multiple generations of stars present in the disks of spiral galaxies due to their extended star formation histories (SFH) and the presence of significant dust extinction complicate the interpretation of Lick indices as degeneracies between different populations are introduced (e.g., [50]). The two approaches that are typically used to overcome this problem are the fitting of linear combinations of single stellar populations (SSP, e.g., [51, 52]) and the use of parametrized SFHs to construct composite stellar populations (e.g., constant and exponentially declining or increasing SFHs, a.k.a. $\pm\tau$ models). After conducting detailed simulations the authors recommend and adopt the latter approach as linear combinations of SSPs are found to perform poorly in the young age and low metallicity regimes. It is worth noting though that these problems, which arise from degeneracies among the large number of parameters inherit to the linear combination of many SSP templates, can be overcome to some extent by using regularization schemes that limit the solutions to conform with certain imposed conditions (e.g., smoothness of the SFH, [52]).

Three of the six disks for which age measurements could be conducted beyond the profile break in Yoachim et al. [20] show evidence for a significant contribution from stellar migration to the stellar populations beyond the truncation radius. This is seen as an upbend in the stellar age radial profile at a position close to the surface brightness profile break and a transition from increasing SFHs to decreasing SFHs towards large radii. In these three objects the changes in stellar population parameters are also clearly seen in the radial profiles of the D4000 and Balmer absorption indices. Two of these objects also show a decrease in stellar metallicity which is correlated with the increase in age at large radii. No significant change in metallicity is seen for the third system. The authors warn that their metallicity measurements are "rather crude" as their SFHs do not include a chemical enrichment model.

For the other three objects with good stellar population measurements at large radii the authors find "*little or no sign of the outer disk being formed through stellar migration despite having some of the strongest profile breaks*". Although these objects do not show up bending age profiles at large radii like the ones discussed above, they do show a flattening in the stellar age profile towards large radii. An interesting result, although based on low number statistics, is that the objects with the weakest profile breaks show the strongest evidence for radial migration and the objects with the strongest profile breaks show young stars most likely formed in situ in the outer disk. This is consistent with stellar migration as objects which have experienced the most migration should have the weakest profile breaks. Given the wide range of stellar population behaviors found beyond the profile breaks of these galaxies the authors conclude that "*while radial migration can contribute to the stellar populations beyond the break, it appears that more than one mechanism is required to explain all of our observed stellar profile breaks*".

### 3.2. Galaxy Dynamics: Dark Matter Haloes and Super Massive Black Holes.

Dark matter is a fundamental mass component in galaxies and drives both the dynamics and evolution of these systems. Similarly SMBHs, which appear to live in the center of all massive galaxies in the universe, seem to be closely linked to the evolution of the central stellar component of galaxies as are evidenced by the tight correlations seen between black hole (BH) mass ($M_{BH}$) and the stellar velocity dispersion, stellar mass, and luminosity of central stellar spheroids ([53–55] and references therein). It is not well established if this link is caused by the effects of AGN feedback following gas accretion or simply by the averaging effects of subsequent mergers (e.g., [56, 57]). Both dark matter and SMBHs affect the dynamics of stars and gas in galaxies in a way that can be significantly detected using modern spectroscopic techniques. The Mitchell Spectrograph has been systematically used to measure the stellar and gaseous dynamics of nearby galaxies in order to put constraints on the shape of the DM halos in that they inhabit and the masses of their central SMBHs.

A problem that has received a lot of attention in the last decade is the apparent discrepancy between the cuspy DM halo profiles predicted by cosmological N-body simulations (e.g., NFW, [58]) and observations of low surface brightness and late-type dwarf galaxies which suggest the presence of cored DM halos (e.g., [59–61]). In Adams et al. [13] the authors use the 2400 lines mm$^{-1}$ grating in the Mitchell Spectrograph which provides a velocity resolution of 40–60 km s$^{-1}$ to measure the stellar and ionized gas kinematics of the late-type dwarf galaxy NGC 2976. This galaxy was one of the cleanest examples of a cored DM halo based on ionized gas kinematics [59]. The goal of Adams et al. [13] study is to constrain the shape of the DM halo in NGC 2976 by using a noncollisional dynamical tracer of the potential (i.e., the stellar component) and compare the results to those obtained using gas kinematics.

Using a mass model based on a multi-Gaussian expansion fit to the $R$-band and HI 21 cm maps to account for the stellar and atomic gas components and conducting an anisotropic Jeans modeling of the second moment velocity field $V_{rms} = \sqrt{V_{los}^2 + \sigma^2}$, [62] the authors put simultaneous constraints on the inner slope of the DM halo profile $\alpha$ and the $R$-band stellar mass-to-light ratio $\Upsilon_{*,R}$. The results from this fitting are shown in Figure 3. The authors find that if both the profile slope and the mass-to-light ratio are fitted freely, both parameters are highly correlated and no significant statistical distinction can be made between the cusp and core scenarios (Figure 3(a)). In this case the best-fit value for $\Upsilon_{*,R}$ is 3.49 which is largely inconsistent with the $\Upsilon_{*,R} = 1.1 \pm 0.8$ value derived from synthetic stellar population fits to both the integrated spectrum and the optical through near-IR photometric SED of the galaxy. Repeating the fit adopting the value and uncertainty of the mass-to-light ratio derived from stellar populations yields the constraints shown in Figure 3(b). If a realistic value of the stellar mass-to-light ratio is adopted, a cuspy DM halo ($\alpha = 1$) profile is consistent with the observed stellar dynamics while a cored profile ($\alpha = 0$) is rejected at 2.2$\sigma$ significance. This is at odds with the results of Simon et al. [59] where the ionized gas velocity field measured from H$\alpha$ is better fit by a cored model.

In Adams et al. [13] the authors also measure the ionized gas velocity field using the [OII]$\lambda$3727 doublet. Modeling the gas dynamics with a harmonic decomposition method [63] similar to the one used by Simon et al. [59] yields results that are consistent with that study. That is, a cored DM halo is preferred. On the other hand using a tilted-ring model with a variable position angle to fit the velocity field [64] excludes a cored model and is compatible with the results from the stellar kinematics analysis. These discrepancies highlight the challenges faced by measurements of the radial mass profile of DM halos, where results can depend on both the adopted dynamical tracer of the gravitational potential and the methodology used to model the observed dynamics. The authors conclude that *"the analysis of this first galaxy shows promising evidence that dark matter halos in late-type dwarfs may in fact be more consistent with cuspy dark matter distributions than earlier work has claimed"*.

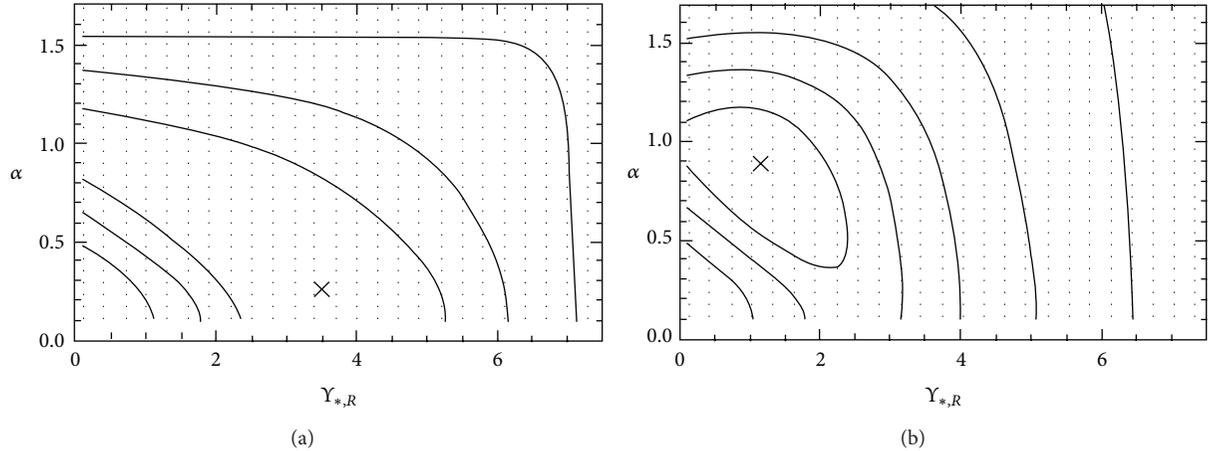

FIGURE 3: Confidence intervals for the inner slope of the DM halo ($\alpha$) and the $R$-band mass-to-light ratio ($\Upsilon_{*,R}$) for NGC 2976. The left panel shows the result of leaving both parameters free while fitting. The right panel shows the same fit but constraining $\Upsilon_{*,R}$ to lie within the limits allowed by a synthetic stellar population fit to the integrated optical spectra and SED of the galaxy. Taken from Adams et al. [13].

The Mitchell Spectrograph has also been used to study the properties of DM halos in massive early type galaxies. Murphy et al. [12] measures stellar kinematics in the second-rank galaxy of the Virgo cluster M87 out to $238''$ (~$2\,R_e$) by tiling five IFU pointing. The authors complement the dataset with higher spatial resolution IFU spectroscopy from SAURON [65] in the inner $13''$ of the galaxy and globular cluster (GC) kinematics out to $540''$ (~$5\,R_e$). By fitting this extended kinematical information with axisymmetric orbit-based dynamical models [66, 67] the authors constrain the mass and shape of the dark matter halo in M87 as well as the $V$-band stellar mass-to-light ratio ($\Upsilon_{*,V}$) and the behavior of the stellar velocity ellipsoid (SVE) as a function of radius for this system. A central SMBH with a mass of $6.6 \times 10^9\,M_\odot$ is included in the dynamical models ([8]; see below).

The authors find that the dynamics of M87 are best described by a logarithmic DM halo (i.e., a flat core in the center that transitions to a $\rho \propto r^{-2}$ profile at radius $r_c$), which is statistically preferred over a classical cuspy NFW profile. Models without a DM halo are excluded at high significance. Out to $5\,R_e$ the best-fit logarithmic halo implies a total enclosed mass in the system of $5.7^{+1.3}_{-0.9} \times 10^{12}\,M_\odot$ making M87 one of the most massive galaxies in the local universe, with dark matter accounting for 17%, 49%, and 85% of the total mass at 1, 2, and $5\,R_e$, respectively. The stellar mass-to-light ratio is constrained to be $\Upsilon_{*,V} = 9.1 \pm 0.2$. Interestingly the inclusion of the GC kinematics reaching out to large radii has a significant impact on the derived value of $\Upsilon_{*,V}$ yielding much higher values if this data is ignored. This is because the kinematics at large radii strongly influence the measurement of the total enclosed mass in the DM halo so any mass not assigned to the dark matter component artificially boosts the mass-to-light ratio of the stars.

The addition of the Mitchell Spectrograph data extends the stellar kinematics in M87 to larger radii by about an order of magnitude compared to what had been achieved with previous IFU and long-slit datasets (e.g., [68]). This has a significant impact on the derived values of the circular velocity and core radius of the DM halo and highly reduces the covariance between these two parameters seen in previous datasets covering only the central parts of the galaxy. This, together with the impact of the GC kinematics on the derived mass-to-light ratio discussed above, highlights the importance of measuring kinematics out to large radii when studying the dynamics of nearby galaxies. This is an area in which wide field of view IFU spectroscopy is having a tremendous impact.

Finally, in Murphy et al. [12] the orbital structure in the best-fit dynamical model implies a drastic change in the anisotropy of the SVE at around $0.5\,R_e$. Within this radius stars show a significant amount of radial-to-tangential anisotropy ($\sigma_r/\sigma_t \sim 1.1$) while outside this radius the orbits become mildly tangentially anisotropic ($\sigma_r/\sigma_t \lesssim 1$). Although the authors do not explore the origin of this behavior they state that it can be related to the formation mechanism that gave rise to this system.

An accompanying paper to Murphy et al. [12] is that by Gebhardt et al. [8] which discusses the mass of the SMBH in the center of M87. Here the authors add adaptive optics (AO) corrected Gemini NIFS integral field spectroscopy to the SAURON+Mitchell Spectrograph dataset discussed above. The AO data provides spatially resolved kinematics for the central $2''$ (170 pc) of the galaxy at $0.08''$ (7 pc) FWHM spatial resolution and therefore resolves the region where the kinematic influence of the SMBH is important.

The degeneracy discussed above between the DM halo mass and the stellar mass-to-light ratio has an impact on the determination of SMBH masses when the sphere of influence of the BH is not resolved. Neglecting or underestimating the dark halo artificially increases the mass-to-light ratio in the stellar component which translates into underestimated BH masses in the best-fit dynamical models (factors of two are common, e.g., [68]). Therefore if the relatively expensive high spatial resolution data (HST or AO) is not available, one can only expect to be able to measure unbiased BH masses if the DM halo is well constrained by kinematic measurements at large radii.

After removing any contamination from the AGN and the jet in the nucleus of M87 and doing a detailed modeling of the AO PSF in the NIFS data Gebhardt et al. [8] measure a SMBH mass of $6.6 \pm 0.4 \times 10^9$. The previously reported covariance between BH mass, mass-to-light ratio, and dark halo parameters seen in Gebhardt and Thomas [68] disappears once the AO data is included. That is, directly resolving the sphere of influence of the BH yields unbiased BH masses that are independent of the inclusion of the dark halo. The authors further test this by excluding the large radii kinematics ($>100''$) from their modeling finding a modest 2% (0.5$\sigma$) drop in the BH mass. The similarity between the measured BH mass and its uncertainty in this study and the $6.4 \pm 0.5 \times 10^9 M_\odot$ BH mass measured by Gebhardt and Thomas [68] support the idea that unbiased BH masses can be accurately measured if the DM halo is properly modeled and constrained by large scale kinematic measurements which wide field integral field spectrographs like the Mitchell Spectrograph provide. The authors conclude that *"if one has sufficient signal to noise and two-dimensional coverage (e.g., SAURON or VIRUS-P), then it should be possible to measure a BH mass robustly. Thus, it is not necessarily required to resolve the region influenced by the BH. Being able to use data that do not resolve well the BHs influence on the kinematics allows us to study BHs that are either distant or of low mass. Both of these regimes are important for understanding the physical nature of the BH correlations with the host galaxy"*.

A similar study is that by McConnell et al. [10] who also use AO corrected IFU spectroscopy from Gemini+NIFS and Keck+OSIRIS in combination with Mitchell Spectrograph data at large radii to measure the SMBH mass in four brightest cluster galaxies (BCGs) in nearby clusters. Using the same dynamical modeling techniques as Gebhardt et al. [8] they measure SMBH masses of $2.1^{+1.6}_{-1.6} \times 10^{10} M_\odot$ for NGC 4889, $9.7^{+3.0}_{-2.5} \times 10^9 M_\odot$ for NGC 3842, and $1.3^{+0.5}_{-0.4} \times 10^9 M_\odot$ for NGC 7768, while putting an upper limit of $<9.0 \times 10^9 M_\odot$ on the central BH mass in NGC 2832. The results for two of these objects (NGC3842 in Abell 1367 and NGC 4889 in the Coma Cluster) were previously reported in McConnell et al. [9] as the two highest BH masses ever measured in the local universe.

Interestingly the measured BH masses in all these objects are systematically higher than those predicted by even the steepest versions of the $M_{\rm BH} - \sigma$ relation found in the literature. For NGC 4889 and NGC 3842 the deviations are in excess of the intrinsic scatter in the relation. A similar trend is seen when comparing the measured values to predictions from the $M_{\rm BH} - L$ (i.e., black hole versus bulge luminosity) relation and from less well established relations between BH mass and DM halo mass or circular velocity. The authors state that the data suggests a steepening in the $M_{\rm BH} - \sigma$ relation at the high mass end. Such a steepening is expected if the most massive elliptical galaxies form by means of "dry" (i.e., gas poor) mergers of early-type systems. Furthermore, both McConnell et al. [10] and Gebhardt et al. [8] observe significant tangential anisotropy in the orbital structure in the centers of these galaxies over spatial scales comparable to their photometric cores. A depletion of radial orbits in the centers of these systems is consistent with a scenario in which following a merger a binary BH preferentially ejects stars on radial orbits from the central regions.

In McConnell et al. [11] the authors extend the dynamical modeling on NGC 4889, NGC 3842, and NGC 7768 to allow the inclusion of a spatial gradient in the stellar mass-to-light ratio $\Upsilon$ in order to study the systematic effects that this quantity has on the derived BH masses. It is standard practice in this type of dynamical modeling to assume a constant $\Upsilon$ across the galaxy therefore equating the stellar surface density to the surface brightness in a given band. On the other hand, as we mentioned above when discussing the stellar population work of Greene et al. [21], many of these systems show gradients in broad-band colors and absorption line indexes which are caused by underlying changes in the metallicity and star formation history of the stellar populations. McConnell et al. [11] modify the code by Gebhardt et al. [53] in order to include a radial gradient in $\Upsilon$ which they parameterize as $\alpha = d \log(\Upsilon)/d \log(r)$ and rerun the fitting for a grid of values in the $-0.2 < \alpha < 0.1$ range. This range covers the values allowed by current constraints on the broad-band color and stellar population gradients of these objects which support a slightly negative value for $\alpha$. For example, stellar population modeling of the Mitchell Spectrograph data in NGC 3842 implies $\alpha = -0.13 \pm 0.30$.

The authors find a ∼25% decrease in the BH mass as $\alpha$ is decreased from 0 to $-0.2$. For two of the three galaxies this difference is comparable to the statistical uncertainties in the measured BH mass so the effect is significant. Models with positive $\alpha$ are mildly disfavored by the data, which is consistent with the negative color gradients of the three objects, and yield higher BH masses. It is seen that introducing a gradient with a negative value moves more stellar mass in the dynamical models to smaller radii, therefore requiring lower BH masses to explain the kinematics. This also translates into slightly more massive DM halos necessary to compensate for the lower stellar mass at large radii. This effect might prove important not only for individual BH mass determinations in systems with strong stellar population gradients but also for determining the slope of BH mass scaling relations. The authors state that *"A systematic variation in $\alpha$ with galaxy stellar mass or velocity dispersion can have an impact on the slopes of the $M_{\rm BH} - M_{\rm bulge}$ and $M_{\rm BH} - \sigma$ scaling relations, which are currently determined assuming $\alpha = 0$"*.

*3.3. The ISM of Nearby Star Forming Galaxies.* Nebular emission lines at optical wavelengths are important coolants for ionized gas in HII regions and are therefore typically bright in star forming galaxies. The observed brightness and line ratios of these transitions have been used extensively in the literature to determine important quantities like metallicity, ionization parameter, electron density, dust extinction, and the SFR ([69] and references therein). Furthermore, AGN activity, shocks, and diffuse ionized gas in the ISM can also contribute significantly to the nebulae emission line spectrum of galaxies. The Mitchell Spectrograph has been widely used to study the star formation activity and the properties of ionized gas in nearby star forming galaxies.

A fundamental aspect of galaxy evolution is the formation of stars from cold gas in the ISM. Star formation gives rise to galaxies and drives their evolution by building up their stellar mass, injecting energy and metals into the ISM, and consuming their gas reservoirs. Understanding what sets the SFR in the different environments present within galaxies is therefore of great importance. In Blanc et al. [17] the authors use data gathered as part of the VIRUS-P Exploration of Nearby Galaxies (VENGA, Blanc et al. [29], Section 3.4.1) to study the spatially resolved star formation law (SFL, a.k.a. the Schmidt-Kennicutt Law) in the inner disk of the nearby spiral galaxy NGC 5194 (a.k.a. M 51a). The SFL is the relation between either the atomic, molecular, or total gas surface densities ($\Sigma_{HI}$, $\Sigma_{H2}$, $\Sigma_{HI+H2}$) and the SFR surface density ($\Sigma_{SFR}$). It has been long known that when integrated measurements of these quantities are made over disk galaxies and starburst nuclear regions in nearby galaxies the total gas surface density correlates with the SFR surface density with relatively small scatter and a slope $N \simeq 1.5$ over several orders of magnitude in both quantities [70]. Spatially resolved measurements of the SFL at sub-kpc scales have shown that it is the molecular gas component of the ISM that drives this correlation while the atomic gas surface density and the SFR surface density are largely uncorrelated on small scales ($10^{2-3}$ kpc, e.g. [71, 72]).

Different authors measure different slopes for the spatially resolved molecular SFL with some favoring a value close to $N \sim 1.5$ (similar to that of the integrated SFL, e.g., [71, 73]) while others find slopes closer to unity or slightly lower (e.g., [17, 72, 74, 75]). Determining the slope of this correlation is important as different theoretical models attempting to explain the process of star formation on galactic scales make different predictions regarding its value (e.g., [76–78]). In Blanc et al. [17] the authors combine the Mitchell Spectrograph data with CO (1-0) and HI 21 cm maps from the literature to study the atomic and molecular SFL at ∼200 pc scales across the inner disk of the galaxy. They use the Balmer decrement to calculate nebular dust extinctions and correct the observed H$\alpha$ fluxes in order to measure $\Sigma_{SFR}$. A major advantage of using integral field spectroscopy comes from the ability to use different emission line ratio diagnostics to identify regions contaminated by AGN photoionization and shocks and to separate the contribution to the H$\alpha$ flux from extraplanar diffuse ionized gas and localized star formation in the disk of the galaxy (i.e., HII regions).

The authors find that the level of discrepancy seen in the slope of the molecular SFL in the literature can be partly attributed to systematic biases associated with the different linear regression methods used to fit the correlation. They propose a new Monte Carlo based method for fitting SFL which includes the intrinsic scatter in the relation as a free parameter and is unbiased. This method has been used later by Leroy et al. [79] who provide a detailed analysis of its robustness. Using this new fitting method Blanc et al. [17] find no correlation between $\Sigma_{SFR}$ and $\Sigma_{HI}$, in agreement with previous studies, a clear correlations between $\Sigma_{SFR}$ and $\Sigma_{H2}$ with a slope $N = 0.82$, and a typical gas depletion timescale $\tau = \Sigma_{HI+H_2}/\Sigma_{SFR} \approx 2$ Gyr (see Figure 4). At ∼200 pc resolution the molecular SFL shows an intrinsic scatter $\epsilon = 0.43 \pm 0.02$ dex in good agreement with the measurements by Schruba et al. [80] and Leroy et al. [79]. These results reject a superlinear molecular gas SFL in NGC 5194 at high significance in agreement with the work of Bigiel et al. [46]. This has been latter confirmed using Bayesian inference by Shetty and Ostriker [81]. In Blanc et al. [82] the authors fit the data of Kennicutt et al. [71] using the method proposed by Blanc et al. [17] and derive a best-fit slope of $N = 1.0$, in contrast with the $N = 1.56$ value derived by the original study. It is currently thought that the steeper slope seen for the integrated SFL arises from a combination of two effects. First, a drop of the molecular-to-atomic gas fraction at low densities which steepens the correlation between the SFR and the total gas surface densities and a true steepening of the small scale molecular SFL in the high surface density regime of starbursts caused by a physical transition in the properties of GMCs in an overpressurized ISM (e.g., [76]). Under the normal conditions typical of the disks of spiral galaxies the relation between $\Sigma_{H2}$ and $\Sigma_{SFR}$ on $10^{2-3}$ pc scales is roughly linear. Studies such as Blanc et al. [17] showcase the power of combining optical IFU spectroscopy with observations at other wavelengths (e.g., radio, mm) to obtain a complete picture of the ionized, atomic, and molecular ISM and its relation to the stellar component of galaxies.

A major observational challenge faced by star formation and ISM studies in general is the proper characterization of molecular gas. Measuring and studying molecular gas in galaxies are fundamental to understand star formation and the physical processes setting the balance between the different phases of the ISM. Molecular hydrogen ($H_2$) amounts for the bulk of the mass in molecules in the universe, but its transitions are rarely excited at the typically cold temperatures (∼10 K) of gas inside giant molecular clouds (GMCs). To overcome this observational difficulty, the second most abundant molecule in GMCs, the carbon monoxide (CO) molecule, is typically used as a proxy for estimating the total mass in $H_2$. Rotational transitions of CO observed at millimeter (mm) wavelengths can be easily excited under the typical density and temperature conditions in GMCs and are therefore bright enough to be detectable. Using CO emission to estimate the $H_2$ mass requires knowledge of the CO intensity to $H_2$ column density conversion factor $X_{CO}$ and, while evidence exists that this factor changes in different environments, it is common usage in the literature to assume it has a constant value ([85], and references within).

Blanc et al. [16] use Mitchell Spectrograph observations of the nearby face-on grand design spiral NGC 628 (also taken as part of VENGA; see Section 3.4.1) in combination with CO (1-0) and CO (2-1) maps from different sources as well as *Spitzer* 24 $\mu$ and *GALEX* far-UV data to look for changes in the CO-to-H$_2$ conversion factor $X_{CO}$ across the disk of this galaxy. Their method consists in inverting the molecular gas SFL to estimate the H$_2$ surface density from the SFR surface density. By comparing this estimate to the observed CO surface brightness they measure the radial $X_{CO}$ profile in NGC 628.

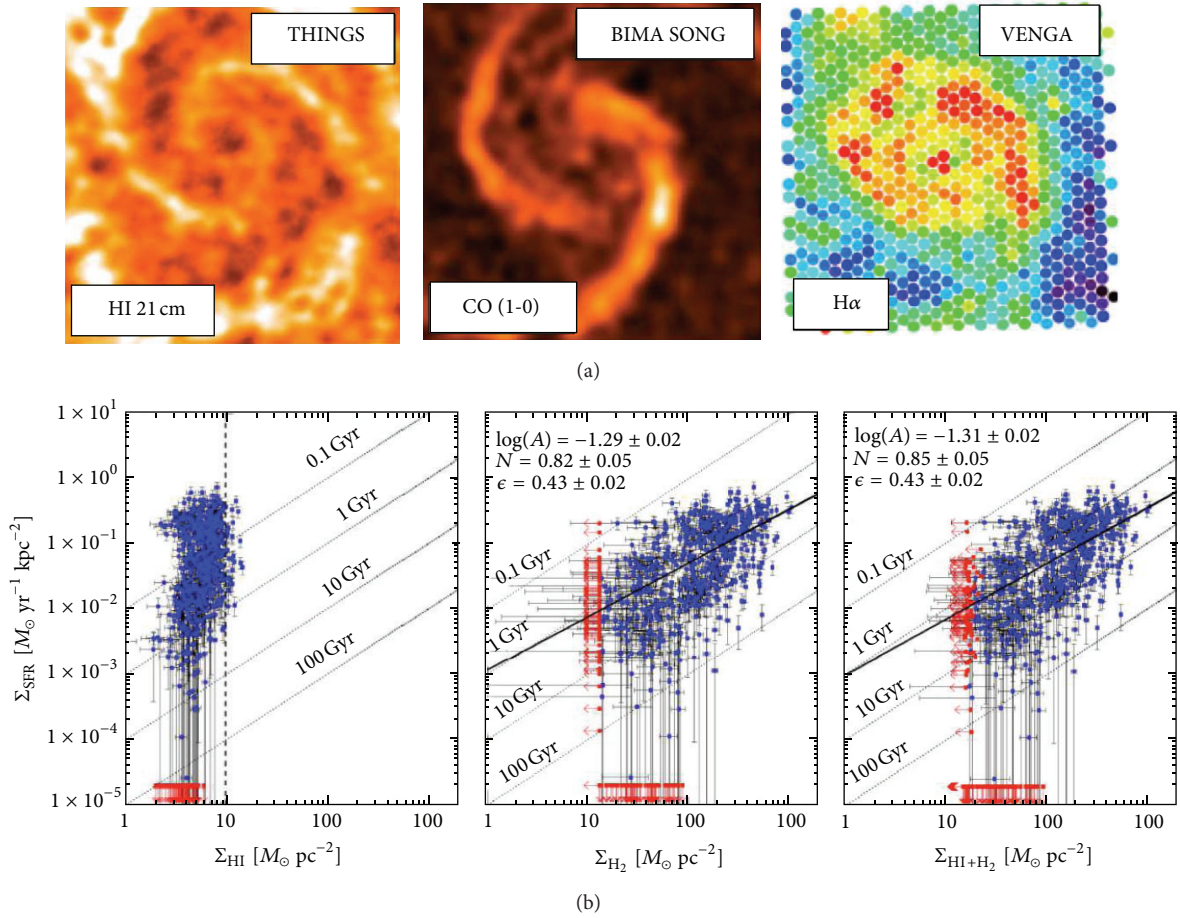

Figure 4: (a) Maps of the HI 21 cm, CO (1-0), and H$\alpha$ emission (from left to right) from the THINGS [83], BIMA-SONG [84], and VENGA [29] surveys used in Blanc et al. [17] to study the spatially resolved SFL in the central 4 kpc of NGC 5194. (b) The HI, H2, and total SFL as measured by Blanc et al. [17]. Each blue dot corresponds to a single Mitchell Spectrograph fiber. Upper limits are shown in red and dotted lines show lines of constant depletion time. The best-fit relations are also shown as well as the corresponding best-fit parameters.

This study finds a drop by a factor of 2 in $X_{CO}$ from $R \sim 7$ kpc to the center of the galaxy, implying brighter CO emission at a given H$_2$ surface density in the central regions. Using line ratio diagnostics from the Mitchell Spectrograph data they can measure how the metallicity and the ionization parameter change across the disk of the galaxy and study how changes in $X_{CO}$ follow changes in these quantities. By comparing to theoretical models for CO radiative transfer in the ISM [86, 87] they show that the observed gradient in $X_{CO}$ can be explained by a combination of decreasing metallicity, and decreasing $\Sigma_{H_2}$ with radius. Photoelectric heating from the local UV radiation field appears to contribute to the decrease

$X_{CO}$ in higher density regions. The authors state that "*Our results show that galactic environment plays an important role at setting the physical conditions in star forming regions, in particular the chemistry of carbon in molecular complexes, and the radiative transfer of CO emission. We caution against adopting a single $X_{CO}$ value when large changes in gas surface density or metallicity are present*".

The Mitchell Spectrograph has also been used to study nebular chemical abundances in nearby galaxies. Integral field spectroscopy allows one to measure gas-phase metallicity gradients in spiral galaxies in an efficient manner, as spectra for multiple HII regions can be obtained in a single IFU pointing. Robertson et al. [88] take advantage of this fact and use the Mitchell Spectrograph to study how the gas-phase metallicity of cluster spiral galaxies depends on HI deficiency. As infall into the intra-cluster medium can remove gas from galaxies by means of ram pressure and tidal stripping and cutoff of the accretion of low metallicity IGM gas into these systems (strangulation), one can expect the environment to affect the chemical enrichment of galaxies by regulating the amount of gas available for star formation and mixing enriched and pristine gas in the ISM.

In Robertson et al. [88] the authors observe six spiral galaxies in the Perseus I cluster with the Mitchell Spectrograph and measure their metallicity gradients using the R23 strong line method of Zaritsky et al. [89]. In about 1 hr of effective exposure time per dither the authors obtain reliable measurements of H$\beta$ and the [OII] and [OIII] doublets for about a dozen HII regions in each galaxy. At the distance of Pegasus I the targets easily fit in the Mitchell Spectrograph field of view. They estimate a characteristic metallicity for each galaxy from the radial metallicity profile and compare this to the HI deficiency parameter (DEF) measured by Levy et al. [90] which quantifies the deviation in HI content with respect to the average expectation value given a galaxy luminosity and Hubble type. A clear correlation between metallicity and DEF is observed with more HI deficient systems showing higher chemical abundances. This is consistent with previous results based on Virgo cluster spirals by Skillman et al. [91].

Robertson et al. [88] test the possibility that the observed correlation is driven by a dependance of DEF with stellar mass combined via the well established mass-metallicity relation of galaxies [92]. They find that for the Pegasus I and Virgo samples DEF is uncorrelated with stellar mass and that using a differential in metallicity with respect to the expectation from the mass-metallicity relation instead of the metallicity absolute value makes the correlation with DEF even stronger. Therefore they conclude that HI deficiency (either by gas removal or strangulation) has a significant effect on the level of chemical enrichment in cluster galaxies in a way that is independent of the stellar mass. The authors compare their results for cluster spirals to a small sample of field galaxies from Zaritsky et al. [89] finding lack of correlation between metallicity and DEF for the latter. This leads them to interpret the observed effect as something particular of dense environments, although a recent extension of their work to a larger and better selected sample of field galaxies shows that the relation between HI deficiency and metallicity is also present in the field and shows a similar slope as that seen in clusters [93].

The Mitchell Spectrograph has also been used recently to study the properties of blue compact galaxies by Cairós et al. [15]. Understanding these type of objects is of great interest as the most massive galaxies in this class are the closest local universe analogs to the high redshift star forming galaxies typically selected using the Lyman break and Ly$\alpha$ techniques. Therefore they offer a window to the conditions under which star formation occurs at high redshift but at a distance at which physical processes can be studied in a spatially resolved manner. Cairós et al. [15] observed the five massive blue compact galaxies II Zw 33, Haro I, NGC 4670, Mrk 314, and III Zw 102 with the Mitchell Spectrograph as part of a long term campaign to obtain integral field spectroscopy of blue compact galaxies. The goal of this study is to characterize the morphology and kinematics of both the stellar and ionized gas components in these objects and to use emission line ratio diagnostics to investigate the spatial distribution of gas ionization and dust extinction. The authors also measure the global gas phase metallicity of the these objects from integrated spectra.

While all the objects analyzed in this study show a relatively smooth and regular morphology in the stellar continuum, a large morphological diversity is seen in the ionized gas which displays more complex patterns with some objects revealing filamentary emission and the presence of bubbles and holes. The galaxies display a variety of different ionization patterns as revealed by maps of the [OIII]$\lambda$5007/H$\beta$ ratio. While in II Zw 33 and Mrk 314 high ionization regions follow star forming regions, as expected from ionization from hot OB stars, NGC 4670 also shows peaks in [OIII]/H$\beta$ over two filaments seen in ionized gas emission above and below the plane of the galaxy, likely caused by shock excitation. Haro I and III Zw 102 on the other hand simply show a positive radial gradient in the line ratio consistent with the metallicity gradients typically seen in spiral galaxies and with the fact that they are the two most massive objects in the sample. All systems show significant amounts of dust extinction and a variety of kinematical behaviors are present in the sample from rotating disks to complex velocity fields consistent with the presence of multiple kinematic components.

Finally, an extremely interesting application that takes advantage of the superb sensitivity to low surface brightness emission of the Mitchell Spectrograph is the study of Adams et al. [14]. In this work the authors attempt to detect H$\alpha$ emission arising from the rim of the outer HI disk of two edge-on late type disk galaxies (UGC 1281 and UGC 7321) as it becomes ionized by the metagalactic ultraviolet background (UVB). If the atomic gas spatial distribution is known at the edge of the HI disk, then a direct measurement of the H$\alpha$ flux can be directly linked to the strength of the UVB. The authors use deep HI 21 cm VLA data on these objects to fit an exponential HI distribution in both the radial and vertical directions. This profile is then extrapolated to the expected position of the disk rim given by the UVB induced photoionization front. This extrapolation seems reasonable since both of the systems analyzed by Adams et al. [14] show extremely thin disks with minimal warping and no signs of flaring at large radii, as well as no massive nearby neighbors which could disturb their HI disks. Despite this, the authors also attempt a measurement that is independent of the above extrapolation and which yields very similar results.

After obtaining extremely deep exposures (~15 hours) using two different gratings providing $R = 1288$ for UGC 1281 and $R = 3860$ for UGC 7321 and trying different binning strategies to coadd the spectra of different fibers, the authors do not detect H$\alpha$ to a 5$\sigma$ limit in surface brightness of $6.4 \times 10^{-19}$ erg s$^{-1}$ cm$^{-1}$ arcsec$^{-1}$ in UGC 7321 and $25 \times 10^{-19}$ erg s$^{-1}$ cm$^{-1}$ arcsec$^{-1}$ in UGC 1281. The difference in depth between the two galaxies is mainly driven by the higher spectral resolution grating used to observe UGC 7321 and exemplify how the combination of large fiber size and high spectral resolution is ideal to achieve good low surface brightness sensitivity. These limits translate in upper limits for the UVB induced HI photoionization rate of $\Gamma(z=0) < 1.7 \times 10^{-14}$ s$^{-1}$ for UGC 7321 and $\Gamma(z=0) < 14 \times 10^{-14}$ s$^{-1}$ for UGC 1281. The UGC 7321 is currently the most stringent limit on the strength of the UVB at $z = 0$ and falls below the predictions of current theoretical models.

*3.4. Nearby Galaxies Surveys.* The Mitchell Spectrograph has been used to conduct two surveys of galaxies in the nearby universe. The first of these surveys, the VIRUS-P Exploration of Nearby Galaxies (VENGA, [29]), is focused on the study of normal star forming disk galaxies while the second one, the VIRUS-P Investigation of the eXtreme ENvironments of Starbursts (VIXENS, [18]), is focused on the study of starburst/interacting systems. Here we describe these observing campaigns, their goals, and some recent scientific results that have come out of them.

*3.4.1. VENGA: The VIRUS-P Exploration of Nearby Galaxies.* VENGA uses the Mitchell Spectrograph to map the disks of 30 nearby spiral galaxies which span a wide range in Hubble type, star formation activity, morphology, and inclination. The Mitchell Spectrograph data can be used to produce 2D maps of the star formation rate, dust extinction, electron density, stellar population parameters, the kinematics and chemical abundances of both stars and ionized gas, and other physical quantities derived from the fitting of the stellar spectrum and the measurement of nebular emission lines. The goal of VENGA is to measure relevant physical parameters across different environments within these galaxies, allowing a series of studies on star formation, structure assembly, stellar populations, chemical evolution, galactic feedback, nuclear activity, and the properties of the interstellar medium in massive disk galaxies.

The survey strategy and data processing methods are thoroughly described in Blanc et al. [29]. The VENGA datacubes have 5.6″ FWHM spatial resolution and ~5 Å FWHM spectral resolution, sample the 3600 Å–6800 Å range, and cover large areas typically sampling galaxies out to ~0.7 $R_{25}$. The VENGA sample together with the fields covered by the Mitchell Spectrograph is shown in Figure 5.

VENGA has produced a series of scientific results regarding the properties of the ISM and the process of star formation in nearby spirals. These include the studies on the SFL in NGC 5194 and the $X_{CO}$ gradient in NGC 628 discussed in Section 3.3 [16, 17]. In the near future a series of studies addressing the chemical composition and the kinematics of the ionized gas in nearby spirals are expected to be published.

*3.4.2. VIXENS: VIRUS-P Investigation of the Extreme Environments of Starbursts.* The VIXENS survey [18] uses the Mitchell Spectrograph to map 15 nearby infrared bright ($L_{IR} > 3 \times 10^{10} L_\odot$) starburst/interacting galaxies. The sample covers a wide range of interaction stages, from early (close pairs) to late phases (coalesced systems with multiple nuclei) with mass ratios in the $1/3 < M_1/M_2 < 1/1$ range. All the objects possess a wealth of ancillary data including *Spitzer* 24 μm, GALEX far-UV, and existing $^{12}$CO(1-0) and HI maps. The VIXENS sample is shown in Figure 6. A parallel campaign to map high density molecular gas tracers (e.g., HCN(1-0), HCO$^+$(1-0), and HNC(1-0)) for the VIXENS sample is being conducted with the Nobeyama 45 m and IRAM 30 m telescopes as well as with the CARMA array. VIXENS will explore the relation between the star formation rate and gas content on spatially resolved scales of ~0.2–1 kpc in starburst/interacting galaxies as a function of interaction phase [18, 97].

## 4. VIRUS and HETDEX: A Look into the Future

As mentioned in Section 1 the Hobby-Eberly Telescope Dark Energy Experiment (HETDEX) will use the VIRUS instrument to conduct a blind spectroscopic survey covering an effective area of ~90 degrees$^2$. While the main goal of HETDEX is to detect Ly$\alpha$ emitters at high redshift ($2 < z < 3.5$), large numbers of lower redshift objects, including nearby galaxies and MW stars, will fall in the survey footprint. HETDEX will reach a depth in continuum of $m_{AB} = 22.6$ (21.5, 20.7) at S/N = 3 (5, 10) per fiber per resolution element. VIRUS observes in the 3500–5500 Å range, has a resolving power of $R = 700$, and a spatial resolution given by the 1.5″ diameter fiber size. Given these parameters, $g < 17$ spiral galaxies will typically have S/N > 3 continuum detections per resolution element per fiber out to 2 effective radii and emission line spectra out to at least their optical radius. HETDEX will spatially resolve ~4500 local galaxies down to that limit without any photometric preselection, and an additional ~10000 local galaxies will have spatially resolved (i.e., $D_{25} > 5″$) spectroscopy below that limit $17 < g < 19$ (Niv Drory private communication). At g = 19 HETDEX will still obtain integrated galaxy spectra at S/N = 10 per resolution element in the continuum. These parameters and the lack of photometric pre-selection make HETDEX very competitive and complimentary with ongoing and future surveys like SAMI [98] and MANGA (http://www.sdss3.org/future/manga.php), although science applications will be constrained to the restricted blue wavelength range of VIRUS.

The spatially resolved absorption and emission spectra will provide information on star formation and dust extinction, the ISM chemical abundance and ionization state, and the properties of stellar populations, as well as rotation curves for an unbiased galaxy sample unprecedented in size. A wealth of information regarding a galaxy's formation history is encoded in the spatial distribution (mostly gradients) of these properties within galaxies, so moving beyond single-fiber (SDSS-like) spectra for large samples of spatially resolved galaxies opens a new parameter space for future studies of galaxy formation and evolution.

## 5. Conclusions

The Mitchell Spectrograph has proven to be a unique instrument for studying nearby galaxies. It is characterized by its large FOV and its high sensitivity to low surface brightness emission which make it an extremely efficient tool to observe extended sources. As I have reviewed throughout this article, the instrument has produced a large number of scientific results regarding the properties of nearby galaxies and the physical processes involved in the formation and evolution of their components, including their stars, ISM, dark matter halos, and central SMBHs.

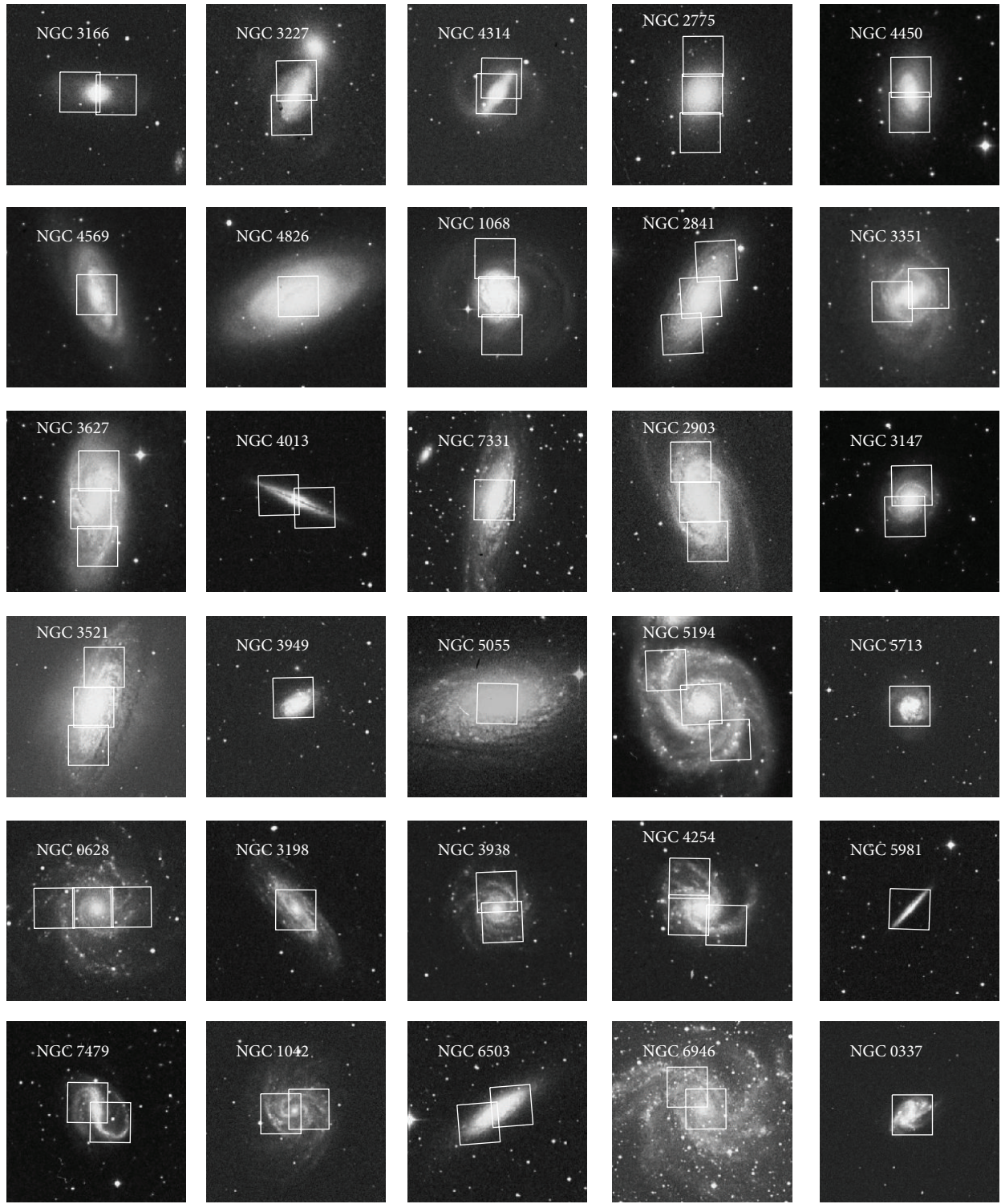

Figure 5: Digital Sky Survey cutouts of the 30 galaxies in the VENGA sample. The targets are ordered by Hubble type (taken from RC3, de Vaucouleurs et al. [94–96]) from earlier to later. White boxes show the VIRUS-P $1.7' \times 1.7'$ pointing observed on each galaxy. Taken from Blanc et al. [29].

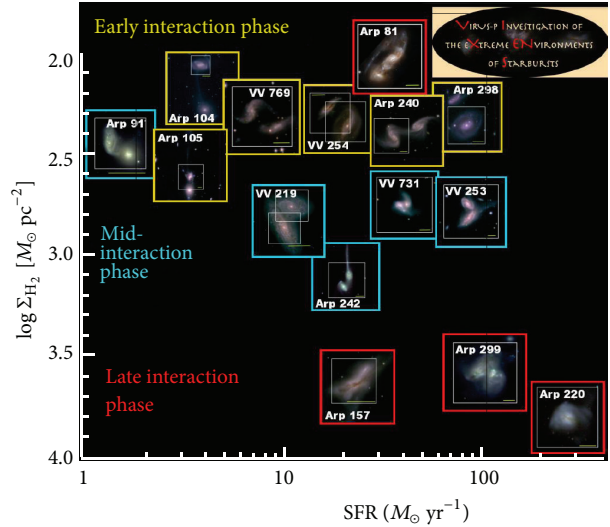

Figure 6: SDSS RGB (i,r,g) images of the VIXENS interacting galaxy sample. White boxes show the Mitchell Spectrograph field of view and the yellow lines indicate a scale of 10 kpc. Taken from Heiderman et al. [18].

The commissioning of the Mitchell Spectrograph on the McDonald 2.7 m telescope has boosted the scientific output of this facility enormously. Wide field integral field spectrographs are extremely powerful scientific instruments for small and medium sized telescopes. The Mitchell Spectrograph and similar instruments in its category have shown during the last decade that relatively modest investments in advanced instrumentation can turn a relatively small telescope into a competitive forefront facility with unique capabilities. Given the difficulties being faced by scientific funding agencies across the world and the current prospects for small optical facilities in the US, this is a very important concept to keep in mind.

Beyond single IFU spectrographs, more ambitious projects like HETDEX and MANGA will soon begin to move the IFU business into the "large data" era. Spatially resolved spectroscopy for tens of thousands of galaxies will soon be available to us. An enormous effort is currently underway to develop the tools necessary to digest and analyze these enormous datasets. While these projects will undoubtedly have an enormous impact on our understanding of galaxies, more detailed studies on smaller numbers of sources for which high spatial resolution, high S/N, full multiwavelength datasets are available will continue to provide a reference point to interpret the statistical results coming out of big surveys. On all fronts integral field spectroscopy is becoming a widely used technique and it is providing an invaluable contribution to the field of galaxy astrophysics.

## Acknowledgments

The construction of the Mitchell Spectrograph (formerly VIRUS-P) was possible thanks to the generous support of the Cynthia & George Mitchell Foundation. Gary Hill and Phillip McQueen at McDonald Observatory lead the design and construction of the spectrograph. The IFU and optical fiber bundle were built by the Leibniz Institute for Astrophysics (AIP). The author acknowledges David Doss and the staff at McDonald Observatory for their invaluable help supporting observations. HETDEX is led by the University of Texas at Austin, McDonald Observatory, and Department of Astronomy with participation from the Universitäts-Sternwarte of the Ludwig-Maximilians-Universität Munchen, the Max-Planck-Institut für Extraterrestriche-Physik (MPE), Astrophysikalisches Institut Potsdam (AIP), Texas A&M University, Pennsylvania State University, and the HET consortium. In addition to Institutional support, HETDEX is funded in part by gifts from Harold C. Simmons, Robert and Annie Graham, The Cynthia and George Mitchell Foundation, Louis and Julia Beecherl, Jim and Charlotte Finley, Bill and Bettye Nowlin, Robert and Fallon Vaughn, Eric Stumberg, and many others and by AFRL under Agreement no. FA9451-04-2-0355, the Texas Norman Hackerman Advanced Research Program under Grants 003658-0005-2006 and 003658-0295-2007, and the National Science Foundation under Grant AST-0926815.

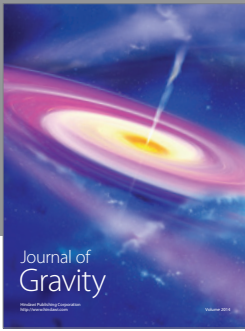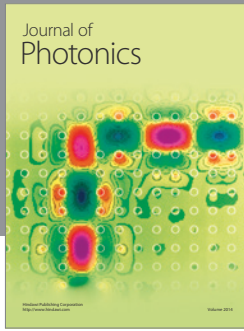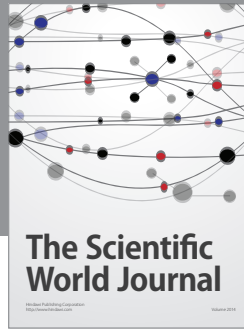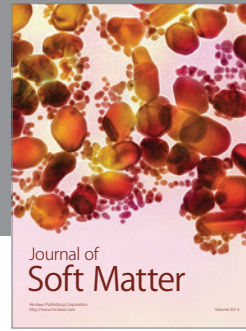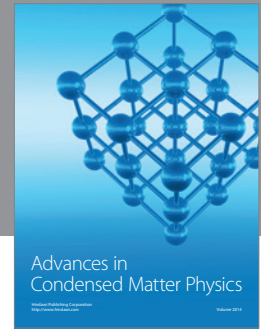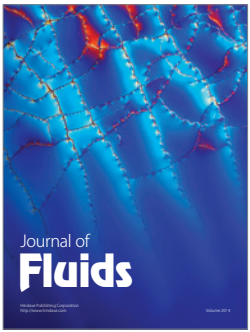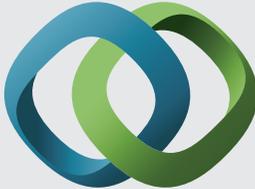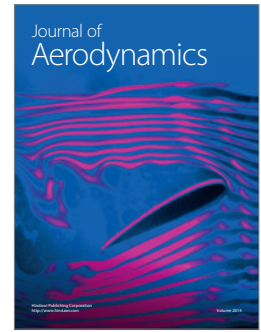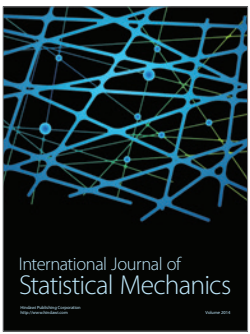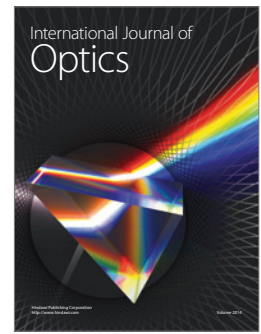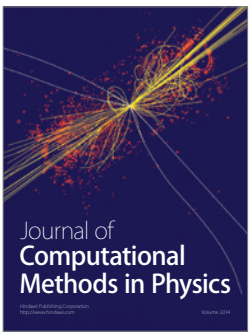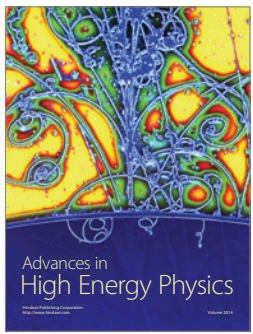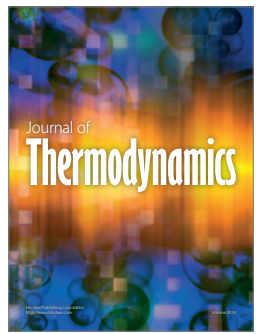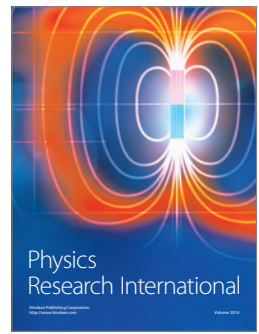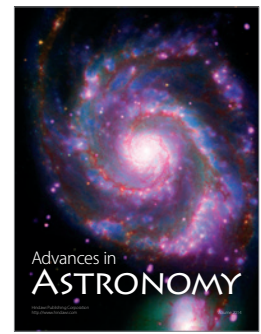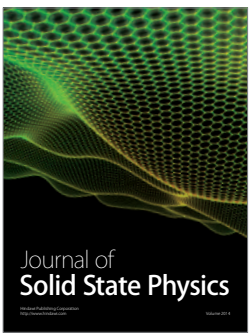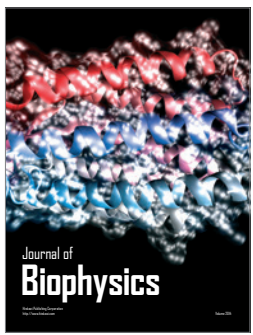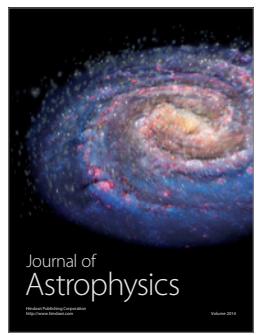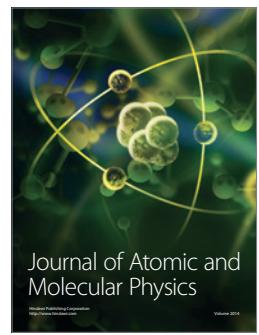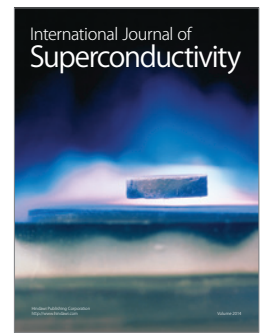